\documentclass[prl,twocolumn,superscriptaddress,floatfix,preprintnumbers,amssymb,amsmath,aps]{revtex4-2}

\usepackage{hyperref}
\usepackage{graphicx}
\usepackage{xcolor}

\begin{document}
\title{Tunable diode effect in a superconducting tunnel junction with biharmonic drive}

\author{David~Scheer}
\email{david.scheer@rwth-aachen.de}
\affiliation{Institute for Quantum Information, RWTH Aachen University, 52056 Aachen, Germany}

\author{Rubén Seoane Souto}
\affiliation{Instituto de Ciencia de Materiales de Madrid, Consejo Superior de Investigaciones Científicas (ICMM-CSIC), 28049, Madrid, Spain}

\author{Fabian~Hassler}
\affiliation{Institute for Quantum Information, RWTH Aachen University, 52056 Aachen, Germany}

\author{Jeroen Danon}
\affiliation{Department of Physics, Norwegian University of Science and Technology, NO-7491 Trondheim, Norway}

\begin{abstract}
A Josephson diode is a superconducting circuit element that enables non-reciprocal transport, allowing a dissipationless supercurrent to preferentially flow in a single direction.
Existing methods for achieving the required symmetry breaking mostly rely on specifically-designed  materials or carefully-engineered circuits composed of multiple Josephson junctions.
Here, we demonstrate that applying a biharmonic drive to a conventional superconducting tunnel-junction induces a diode effect through harmonic mixing processes that shift the supercurrent region.
We show that, in a conventional tunnel junction, unity efficiency is achievable while maintaining a large supercurrent.
Moreover, the relative phase between the two driving tones determines the directionality of the diode, which can be tuned in situ.
\end{abstract}
\maketitle

\textit{Introduction}.---The simultaneous breaking of time-reversal and inversion symmetries in a superconducting system can lead to phenomena such as the anomalous Josephson effect and a non-reciprocal supercurrent response~\cite{Edelstein1989,PhysRevLett.75.2004,Margaris2010,PhysRevB.89.195407,PhysRevB.92.035428,PhysRevB.93.155406,Wakatsuki2017,PhysRevB.98.075430}.
The resulting rectification of supercurrent, known as the `\emph{superconducting diode effect}', has attracted significant attention in recent years~\cite{ando2020observation,PhysRevLett.128.037001,he2022phenomenological,PhysRevX.12.041013,Yuan_PNAS2022,PhysRevLett.128.177001,Legg_PRB2022,Wang_arXiv2022,Karabassov_PRB2022,PhysRevB.108.214520,Steiner_PRL2023,Maiani_PRB2023,Kotetes2024} due to its potential applications in non-dissipative superconductor electronics~\cite{Semenov_IEE2015,Braginski_JSNM2019}.
The superconducting diode effect has been observed in various systems, including superconducting thin films~\cite{Hou_PRL2023}, semiconductor-based Josephson junctions~\cite{Baumgartner_NatNano2022,Lotfizadeh2024}, transition metal dichalcogenides~\cite{Bauriedl_NatCom2022}, twisted multilayer graphene~\cite{Lin_NatPhys2022,Diez-Merida2023}, and topological semimetals~\cite{Pal_NatPhys2022}. It has also been observed in devices where inversion symmetry is intentionally broken, such as those featuring artificial superlattices~\cite{ando2020observation} or arrays of nanoholes~\cite{Lyu_NatPhys2021}. A recent review of the superconducting diode effect can be found in Ref.~\cite{Nadeem_NRM2023}.

Rather than depending on specially designed materials and junctions with inherently broken symmetries, a superconducting diode can also be realized by combining multiple conventional Josephson junctions, which are easily fabricated and well-controlled. This approach includes using superconducting interferometers~\cite{Fominov2022,Souto2022,Ciaccia2023,Valentini2024,Leblanc_arXiv2023}, multi-terminal Josephson junctions~\cite{Chiles_NanoLett2023,Gupta_NatCom2023}, or arrays of Josephson junctions~\cite{Haenel2022} that can even be used to engineer an arbitrary current-to-phase relation (CPR) through hundreds of finely-tuned junctions~\cite{Bozkurt2023}.

It is well-known that the dc transport properties of a Josephson junction can also be modified by applying an external ac current. In quantum metrology, a harmonic drive at frequency $\omega$ is commonly used to create Shapiro steps of a constant voltage corresponding to integer multiples of the frequency $V_n=n\hbar\omega/2e$, where $\hbar$ is the reduced Planck constant and $2e$ is the charge of a Cooper pair~\cite{Shapiro1963,Jeanneret2009}. Besides forming Shapiro steps,  the drive also modifies the width $\Delta I$ of the supercurrent region by renormalizing the critical current of the junction~\cite{Likharev1986}. 

While a single frequency drive does not alter the symmetry properties of the junction, a drive  composed of two frequencies can be used to introduce an asymmetry in the IV-curve.
Biharmonic drives have been extensively analyzed in the context of the ratchet effect, where they have been employed to achieve directed energy transport without an external bias~\cite{Marchesoni1986, Goychuk1998, Flach2000, Ustinov2004, Chacon2007,Machura2012}. 
Moreover, recent experimental findings show that a non-sinusoidal drive can also used to create asymmetric dual Shapiro steps~\cite{Kaap2024} in a phase-slip junction where a potential diode effect was masked by the strong influence of thermal fluctuations present in phase-slip experiments.
While there have been investigations on tuning the parameters of an already existing Josephson diode by a single frequency drive~\cite{Souto2022,Souto2024} and a diode effect in a driven double quantum dot~\cite{Ortega-Taberner2023}, the possibility of a diode effect through a symmetry-breaking drive in a conventional tunnel junction has not yet been discussed.

In this letter, we show that a biharmonic drive can induce a diode effect in a superconducting tunnel junction with an inversion-symmetric CPR by breaking the spatio-temporal symmetries of the equation of motion in the resistively shunted junction-model for overdamped junctions. 
Specifically, we explore how the relative phase of the driving tones can be adjusted to tune the breaking of the additional shift symmetry present in the CPR of a Josephson junction. We show that the biharmonic drive can produce a superconducting diode with unit efficiency in a tunable direction.  We illustrate how the resulting diode effect can be understood as an adiabatic modulation of the bias current in the slow driving frequency limit. Beyond the slow driving regime, we demonstrate that the diode effect results from a cascade of two-tone mixing processes that produce an effective shift of the supercurrent region for small driving amplitudes. Finally, we provide a brief outline of a potential experimental realization of the diode and compare it to other state-of-the-art approaches.

\begin{figure}
    \centering
    \includegraphics[width=\linewidth]{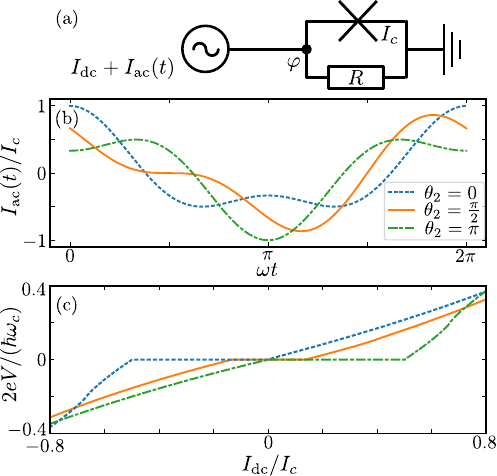}
    \caption{Diode response in the regime of slow driving. (a) Electric-circuit under investigation. (b) Biharmonic driving signals at frequency $\omega/\omega_c=10^{-3}$ with amplitudes $I_2/I_1=1/2$ and $I_1/I_c=2/3$ for different values of the phase $\theta_2$. (c) Resulting IV-curves for different values of $\theta_2$. While none of the signals exhibits a shift-symmetry, the signal with $\theta_2=\pi/2$ still retains a point symmetry which prohibits a diode effect. The other signals reach the critical current in one direction and therefore produce a diode with unit efficiency. By tuning the phase $\theta_2$, the behavior of the junction  continuously interpolates between the  curves presented.}
    \label{fig:slow_driving}
\end{figure}

\textit{Symmetry breaking drive}.---We consider the dynamics of the superconducting phase $\varphi$ across a Josephson junction with a current bias and a shunt resistance $R$. For small enough $R$, the circuit shown in Fig.~\ref{fig:slow_driving}(a) is in the overdamped limit and can be described by
\begin{align}\label{eq:RSJ-model}
    \frac{\hbar}{2eR}\dot\varphi+I(\varphi)=I_{\rm dc}+I_{\rm ac}(t),
\end{align}
where $I(\varphi)$ is the CPR of the Josephson junction and $I_{\rm ac}(t)$ is a periodic driving current with period $T$ and without a dc-component. The dc-voltage across the junction is given by $V=\hbar\langle\dot\varphi\rangle/(2e)$, where $\langle\cdot\rangle$ denotes the time average.

We investigate the conditions under which Eq.~(\ref{eq:RSJ-model}) predicts a symmetric response to the applied bias current $I_{\rm dc}$, \emph{i.e.}, when the existence of a solution $\varphi(t)$ [corresponding to the dc-voltage $\hbar\langle\dot\varphi\rangle/(2e)$] for the bias current $I_{\rm dc}$ implies that the current $-I_{\rm dc}$ must yield the solution $-\varphi(t)$ with opposite voltage $-V$.

For generic weak-link junctions, the CPR exhibits an inversion symmetry $I(-\varphi)=-I(\varphi)$ due to the microscopic time-reversal symmetry \cite{Likharev1979}.
This implies that without an external drive, $I_{\rm ac}=0$, the system cannot exhibit a diode effect:
The simultaneous substitutions $\varphi \mapsto -\varphi$ and $I_{\rm dc} \mapsto -I_{\rm dc}$ leave Eq.~(\ref{eq:RSJ-model}) unchanged, thus guaranteeing solutions with $\pm V$ for $\pm I_{\rm dc}$.
When an external drive is added, $I_{\rm ac}\neq 0$, the symmetry of the IV-curve remains unchanged, as long as the driving function exhibits a shift symmetry $I_{\rm ac}(t)=-I_{\rm ac}(t+\tilde\tau)$ (implying $\tilde\tau = T/2$).
Indeed, in this case the simultaneous substitutions $\varphi \mapsto -\varphi$, $t\mapsto t+\tilde\tau$ and $I_{\rm dc} \mapsto -I_{\rm dc}$ map Eq.~(\ref{eq:RSJ-model}) onto itself, thus still prohibiting a diode effect.
Therefore, the shift symmetry of the drive must be broken to allow for a diode effect.

The simplest way to break the shift symmetry is to use a biharmonic drive
\begin{equation}\label{eq:biharmonic}
    I_{\rm ac}(t)=I_1\cos(\omega t)+I_2\cos(2\omega t+\theta_2),
\end{equation}
with an adjustable phase $\theta_2$. Such drives have been widely analyzed in the context of breaking spatio-temporal symmetries for controlling a ratchet effect \cite{Chacon2007}.
In the context of Josephson junctions, previous investigations were focused on two coupled junctions ~\cite{Machura2012} and on the modified properties of Shapiro steps~\cite{Kanter1972,Wonneberger1983, Likharev1986,Monaco1990} without a dedicated treatment of the effect on the supercurrent region.

In the case of a tunnel junction, the CPR is given by $I(\varphi)=I_c\sin(\varphi)$ with $I_c$ its critical current.
This CPR exhibits an additional shift symmetry $I(\varphi+\pi)=-I(\varphi)$ that further enforces the symmetry of the IV-curve.
With this shift symmetry present, a diode effect is also prohibited if the drive has a point symmetry, $I_{\rm ac}(\tilde\tau -t)=-I_{\rm ac}(t)$, making the joint substitutions $\varphi \mapsto \varphi +\pi$, $t \mapsto \tilde\tau - t$ and $I_{\rm dc} \mapsto -I_{\rm dc}$ a symmetry operation of Eq.~(\ref{eq:RSJ-model}).
The biharmonic drive in Eq.~\eqref{eq:biharmonic} retains such a point symmetry for $\theta_2=\pm\pi/2$ (with $\tilde\tau = T/2$).
For all other phase values the drive is chiral with different amplitudes in both directions $|I_{\rm ac}^+|\neq|I_{\rm ac}^-|$, where $I_{\rm ac}^+$ $(I_{\rm ac}^-)$ is the maximum (minimum) value of the driving current $I_{\rm ac}(t)$.
In Fig.~\ref{fig:slow_driving}(b) we show a comparison between the symmetric drive at $\theta_2=\pi/2$ and the most asymmetric drives at $\theta_2\in\{0,\pi\}$. By this tunable breaking of symmetries, a biharmonic drive can thus produce a diode effect in a symmetric junction. 

\textit{Slow driving limit}.---The effect of an asymmetric drive can be most easily understood in the regime where the driving frequency $\omega$ is much smaller than the typical relaxation timescale of the circuit given by $\omega_c=2eI_cR/\hbar$.  In this regime, the dynamics of the superconducting phase correspond to a massless particle moving in a tilted washboard potential. Due to the lack of inertia, the particle can get stuck in a local minimum with zero average velocity $\langle\dot\varphi\rangle=0$ which corresponds to a vanishing dc-voltage. As explained in~\cite{Souto2024}, the junction will exhibit zero average voltage as long as the combined driving current and bias current do not exceed the critical current.
This leads to a simple condition for the directional critical currents
\begin{align}
    I_c^-=-I_c-I_{\rm ac}^-\leq I_{\rm dc}\leq I_c-I_{\rm ac}^+=I_c^+,
\end{align}
where the resulting diode efficiency is given by
\begin{align}
    \eta=\frac{|I_c^++I_c^-|}{|I_c^+-I_c^-|}=\frac{|I_{\rm ac}^++I_{\rm ac}^-|}{|2I_c-I_{\rm ac}^++I_{\rm ac}^-|}.
\end{align}
A unit diode efficiency ($\eta=1$) occurs, when the larger one of the directional driving amplitudes reaches the critical current ${\rm max}|I_{\rm ac}^{\pm}|=I_c$ since any added dc-bias will overcome the critical current. If the amplitude of the drive is further increased, the circuit exhibits a ratchet effect with a finite voltage at zero bias current that formally corresponds to $\eta>1$.
In Fig.~\ref{fig:slow_driving}(c), we show IV-curves calculated numerically from Eq.~(\ref{eq:RSJ-model}), for a drive at frequency $\omega/\omega_c=10^{-3}$ with amplitudes $I_2/I_1=1/2$ and $I_1/I_c=2/3$. The phase $\theta_2$ can be adapted to achieve unit diode efficiency in either direction as well as no diode effect. Note that the small Shapiro steps produced by the drive are not resolved in the plot. 

\textit{Cascaded two-tone mixing}.---While the slow-driving limit offers a good understanding of the diode effect, it is not well-suited to produce a diode for practical applications. This is due to the fact, that a slowly driven junction will only act as a diode for dc-currents. If the diode is embedded in a circuit with non-zero operating frequencies $\omega_{\rm op}$, it needs to behave as a diode at least up to $\omega_{\rm op}$. In order to produce a diode at a broad range of frequencies, the operating frequencies should be much slower than the relaxation time of the junction and the driving frequency which can be of the order of $2\pi\times10\,$GHz.

In this faster regime, the diode effect can not be understood in the simple adiabatic picture of the drive acting as an additional bias current. Instead, it relies on an effective dc-bias current that is produced by a cascade of harmonic mixing processes. For a Josephson junction with a biharmonic drive, Eq.~(\ref{eq:RSJ-model}) reduces to
\begin{align}
    \dot\varphi+\sin(\varphi)=i_{\rm dc}+{\rm Re}\Bigl[i_1 e^{i\nu \tau}+i_2 e^{i(2\nu \tau+\theta_2)}\Bigr],
\end{align}
with the dimensionless variables  $i_j = I_j/I_c$, $\tau=\omega_c t$, and $\nu=\omega/\omega_c$.

In order to analyze the resulting diode effect from the drive, we consider the boundary points $i_{\rm dc}=\pm1$ of the supercurrent region in the undriven case. We separate the dc-response of the superconducting phase introducing the deviations $\varphi_\pm$ from the phase $\pm \pi/2$ corresponding to the boundary points, \emph{i.e.}, $\varphi= \pm\pi/2+\varphi_\pm$ with  $\langle\dot\varphi_\pm\rangle=0$. At small driving strengths $i_j\ll 1$, we can expand in the small response to the drive $\varphi_\pm\ll 2\pi$ in order to obtain
\begin{align}
    \dot \varphi_\pm=\delta i_{\rm dc}+{\rm Re}\Bigl[i_1 e^{i\nu \tau}+i_2 e^{i(2\nu \tau+\theta_2)}\Bigr]\pm\frac{\varphi_\pm^2}{2}+\mathcal{O}(\varphi_\pm^4);
\end{align}
her we introduced the shift $\delta i_{\rm dc}$ of the bias current to compensate the dc-component of the two-tone mixing term $\varphi_\pm^2$. To leading order, the bias current gets a correction $\delta i_\text{dc}$ due to the two-tone mixing $\pm \frac{1}{2}\langle \varphi^2\rangle$. As the correction has the opposite sign for the two points $I_\text{dc} = \pm I_c$, this simply leads to a decrease of the critical current, see first term of Eq.~\eqref{eq:deltai} below. However, as we illustrate in Fig.~\ref{fig:mixing_cascade}, the mixing term also produces effectively an additional ac-drive, which will inherit the sign of the bias current. By further mixing this correction with the response to the external drive, we obtain an additional dc-component with a sign that does not depend on the bias current $i_{\rm dc}$. This corresponds to a shift of the whole supercurrent region which produces a diode effect.
\begin{figure}[t]
    \centering
    \includegraphics[width=\linewidth]{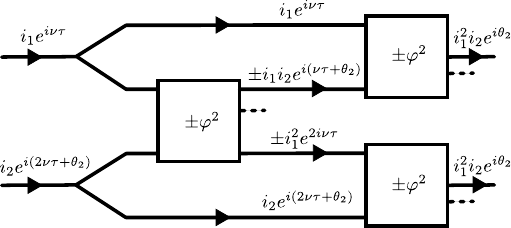}
    \caption{Two-tone mixing cascade. The two tones of the biharmonic driving signal with a phase $\theta_2$ are mixed by the Josephson junction with a sign depending on the direction of the bias current. In addition to higher-frequency components and a phase-independent dc-contribution that do not contribute to the asymmetry at this order (dashed lines), the mixing process also produces additional driving terms at the frequencies of the original drive that inherit the sign of the mixing term. These added driving terms undergo the same mixing process with the original driving terms to produce a dc-current that depends on the phase $\theta_2$ and has the same sign for both directions of the bias current. The cascaded process produces a diode effect.}
    \label{fig:mixing_cascade}
\end{figure}

We can obtain the size of the diode effect by a perturbation expansion in $i_j$. To this end, we set $\varphi_\pm = \varphi_1 + \varphi_2 + \cdots$ with $\varphi_k = \mathcal{O}(i_j^k)$ and obtain
\begin{align}
    \dot\varphi_{1} = {} & {}{\rm Re}\biggl[i_1 e^{i\nu \tau}+i_2 e^{i(2\nu \tau+\theta_2)}\biggr],\\
    \dot\varphi_{2} = {} & {}\delta i_{\rm dc}\pm\frac{\varphi_{1}^2}{2}.
\end{align}
These equations have the solution
\begin{align}
    \varphi_{1} = {} & {}{\rm Im}\biggl[\frac{i_1}{\nu} e^{i\nu \tau}+\frac{i_2}{2\nu} e^{i(2\nu \tau+\theta_2)}\biggr],\\
    \varphi_{2} = {} & {}\pm\frac{1}{4\nu^3}{\rm Im}\biggr[i_1i_2e^{i(\nu \tau+\theta_2)}-\frac{i_1^2}{2}e^{2i\nu \tau}\biggl],
\end{align}
where we neglected the higher frequency terms in $\varphi_{2}$ as they only give sub-leading corrections to the dc-current. The perturbations are  small in the  parameter $i_j/\nu=2e R I_j/(\hbar\omega)\ll 1$ such that this is in fact the condition under which our perturbative approach is valid. Note that this condition does not depend on the critical current $I_c$. The resulting leading order corrections for the renormalized critical current and diode effect are given by
\begin{align}\label{eq:deltai}
     \delta i_{\rm dc}= {} & {} \mp\frac{1}{2}\langle\varphi_{1}^2\rangle\mp\bigl\langle\varphi_{1}\varphi_{2}\bigr
     \rangle.
\end{align}
For the directional critical currents $I_c^{\pm}$, we thus obtain
\begin{align}
    I_c^{\pm}/I_c=\pm\biggl(1-\frac{4i_1^2+i_2^2}{16\nu^2}\biggr)-\frac{3i_1^2i_2}{32\nu^4}\cos(\theta_2),
\end{align}
where the second term causes the asymmetry by a shift of the supercurrent region in a direction determined by the sign of $\cos(\theta_2)$. While the numerical prefactor of the asymmetry term depends on the specifics of the CPR, the proportionality to $i_1^2i_2\cos(\theta_2)/\nu^4$ is the same for a general inversion-symmetric CPR as this results from an expansion about global maximum (minimum) of the CPR. As shown in~\cite{Chacon2007}, this result is expected for biharmonic drives in a generic setting. 
The diode efficiency is
\begin{align}
    \eta=\frac{3i_1^2i_2}{32\nu^4}\bigl|\cos(\theta_2)\bigr|+\mathcal{O}(i_j^5\nu^{-6}).
\end{align}
\begin{figure}[tb]
    \centering
    \includegraphics[width=\linewidth]{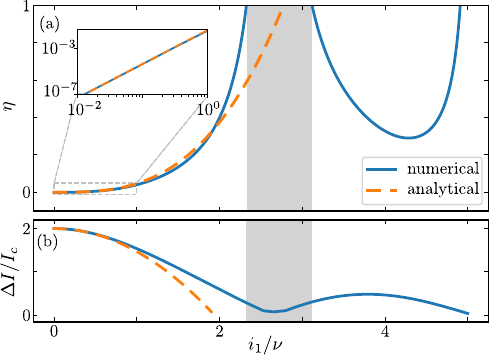}
    \caption{Diode effect at varying driving strength $I_1$ for a biharmonic drive with $\omega=\omega_c$, $I_2/I_1=1/2$, and  $\theta_2=0$. (a) Resulting diode efficiency $\eta$. An efficiency of 1 is reached for the first time around $i_1/\nu = 2eRI_1/(\hbar\omega)\approx 2.3$. An efficiency above 1 corresponds to a ratchet effect at $I_{\rm dc}=0$ (gray shaded region) since the supercurrent region no longer includes the point of zero bias current, at which a finite voltage is produced. In the inset, we show that the analytical results agree with the numerics up to $i_1/\nu \lesssim 1$. (b) Overall width of the supercurrent region. The width of the supercurrent region tends decrease at larger diode efficiency with values of $\Delta I\approx 0.3I_c$ at efficiency 1. }
    \label{fig:diode_efficiency}
\end{figure}

In Fig.~\ref{fig:diode_efficiency}, we show numerical results for the resulting diode efficiency $\eta$ as well as the width of the supercurrent region for a drive with phase $\theta_2=0$, where the driving strength $i_1$ is varied, while the ratio $i_2/i_1=1/2$ is kept at a constant value. For simplicity we choose the driving frequency $\nu=1$. Up to $i_1/\nu\approx1$, the simulated curves show a good agreement with the analytical prediction for both the diode efficiency and the width of the supercurrent region. The maximum diode efficiency of 1 is reached at a driving strength of $i_1/\nu\approx2.3$ and occurs repeatedly for increasing driving strength. When the diode efficiency increases above 1, the junction again exhibits a ratchet effect when the supercurrent region is shifted away from the zero-bias point. While the total width of the supercurrent region decreases with increasing diode efficiency, it is still at a value of $\Delta I\approx 0.3 I_c$ at the point of unit diode efficiency. Therefore, the overall critical current in the `supercurrent direction' of the diode is only decreased by a factor of 3. As in the slow driving case, the direction of the supercurrent region as well as the diode efficiency can be tuned by adjusting the phase $\theta_2$ of the drive.

\textit{Implementation}.---In an experimental setting, the diode can be implemented in the same setup that displays regular Shapiro steps. The main addition is the second harmonic drive at double the target frequency $\omega$ with a stable phase relation to the first drive. In order to remain in the overdamped regime, it is important that both frequency components are slow compared to the plasma frequency of the junction $2\omega\ll\omega_p=\sqrt{2eI_c/(\hbar C)}$, where $C$ is the intrinsic capacitance of the junction. Since the occurrence of the diode effect does not depend on the size of the critical current, $I_c$ can be chosen to be at the order of mA to increase the size of the supercurrent region. While the required external driving presents a drawback in the implementation compared to other state-of-the-art diode designs \cite{Hou_PRL2023,Bozkurt2023}, a driven diode combines the advantages of well-controlled and easily fabricated junctions with the simplicity of a single circuit element. Additionally, it achieves perfect efficiency at high supercurrents, enabling rapid switchability via the driving phase, along with minimal hysteresis in the IV-curve within the overdamped regime.

\textit{Conclusion}.---We investigated the superconducting diode effect that arises in a single tunnel junction induced by a biharmonic drive.  
We demonstrated that this effect originates from a cascade of two-tone mixing processes, resulting in a constant shift of the supercurrent region at small driving strengths.
Additionally, we numerically demonstrated that a biharmonic drive can induce a diode effect with unit efficiency in a reciprocal Josephson junction, with a critical current in the same order of magnitude as in the undriven junction.
We found that the direction of the diode can be switched in situ by varying the relative phase of the driving tones.
Throughout our analysis, we focused on the resulting diode effect within an isolated circuit.
The behavior of the diode as an element in a larger circuit would be question for future research.

We acknowledge fruitful discussions with F.\ Kaap.
This work was supported by the Deutsche Forschungsgemeinschaft (DFG) under Grant No.~HA 7084/6–1, the Spanish Comunidad de Madrid (CM) ``Talento Program'' (Project No. 2022-T1/IND-24070), and the Spanish Ministry of Science, innovation, and Universities under grant PID2022-140552NA-I00.

\end{document}